\documentclass[showpacs,showkeys,byrevtex,twocolumn,pra]{revtex4-1}
\usepackage{epsfig}
\usepackage{amsmath}
\usepackage{amsfonts}
\usepackage{color}
\usepackage{graphicx}
\usepackage{subfigure}
\usepackage{hyperref}
\usepackage{multirow}
\input{Qcircuit}

\def\<{\langle}
\def\>{\rangle}
\def\bra#1{{\langle#1|}}
\def\ket#1{{|#1\rangle}}

\def\bea{\begin {eqnarray*}}
\def\eea{\end {eqnarray*}}

\def\bar{\overline}
\def\*{\star}
\def\({\left(}

\def\){\right)}

\def\2pi{\hbox{$2\pi i$}}

\def\dsl{\raise.15ex\hbox{/}\kern-.57em\partial}
\def\Dsl{\,\raise.15ex\hbox{/}\mkern-.13.5mu D}

\def\<{\langle}
\def\>{\rangle}

\def\CD{{\cal D}}
\def\CE{{\cal E}}

\def\CN{{\cal N}}
\def\CO{{\cal O}}

\def\CU{{\cal U}}

\def\1{{\mathbf{1} }}

\def\2pi{\hbox{$2\pi i$}}

\def\dsl{\raise.15ex\hbox{/}\kern-.57em\partial}
\def\Dsl{\,\raise.15ex\hbox{/}\mkern-.13.5mu D}
\def\beq{\begin {equation}}
\def\eeq{\end {equation}}

\begin{document}

\preprint{ }
\title[ ]{Hiding Quantum Information in the Perfect Code}
\author{Bilal A. Shaw$^{3,4}$}
\email{bilalsha@usc.edu}
\author{Todd A. Brun$^{1,2,3,4}$}
\affiliation{$^{1}$Department of Electrical
Engineering, $^{2}$Department of Physics and Astronomy, $^{3}$Department of Computer Science,
$^{4}$Center for Quantum Information Science and Technology, University of Southern California, Los Angeles, California 90089, USA}
\keywords{quantum steganography, quantum error-correction, stabilizer formalism}

\begin{abstract}
We present and analyze a protocol for quantum steganography where the sender (Alice) encodes her steganographic information into the error syndromes of the perfect (five-qubit) quantum error-correcting code, and sends it to the receiver (Bob) over a depolarizing channel.  Alice and Bob share a classical secret key, and hide quantum information in such a way that to an eavesdropper (Eve) without access to the secret key, the quantum message looks like an innocent codeword with a typical sequence of quantum errors.  We calculate the average rate of key consumption, and show how the protocol improves in performance as information is  spread over multiple codeword blocks.  Alice and Bob utilize different encodings to optimize the average number of steganographic bits that they can send to each other while matching the error statistics of the depolarizing channel.
\end{abstract}
\volumeyear{2010}
\volumenumber{ }
\issuenumber{ }
\eid{ }
\date{\today}
\received{\today}

\revised{}

\accepted{}

\published{}

\pacs{03.67.-a,03.67.Dd,03.67.Hk,03.67.Pp}
\startpage{1}
\endpage{ }
\maketitle

\section{Introduction}
\label{sec:intro}
The art and science of information hiding or steganography has been around for centuries, if not millennia.  The word itself comes from the Greek words \textit{steganos} which means ``covered,'' and \textit{graphia} which means ``writing.''  The art of information hiding dates back to 440 B.C. to the Greeks.  In \textit{The Histories}, Herodotus records two incidents of the use of steganography.  In the first incident Demaratus a Greek king uses a wax tablet to warn the Spartans of an impending attack by the Persian king Xerxes~\cite{herodotus}.  Wax tablets were used as reusable writing surfaces and constructed from wooden bases.  Demaratus scratched the steganographic message on the wood, and then covered it with beeswax.  Once the Spartans received the wax tablet from the courier, all that they needed to do was to melt the wax and read the hidden warning.  In another story Herodotus records how Histiaeus tattoos a secret message on the shaved scalp of his slave, and then waits for the hair to grow back before dispatching him to the Ionian city of Miletus.  The ancient Chinese used a wooden mask with holes cut out at random places to use as a steganographic device.  They would place the wooden block on a blank sheet of paper and after writing the secret message in the holes they would fill in the blanks in the paper with regular text.  The mask acted as a secret key to unlock the hidden message.  In the seventeenth and eighteenth centuries, people hid secret messages in logarithm tables by introducing deliberate errors in the least significant digits.

The term steganography was first used in 1499 by Johannes Trithemius in his \textit{Steganographia} which was one of the first treatises on the use of cryptographic and steganographic techniques~\cite{trithemius}.  The study of modern steganography was initiated by Simmons, who presented it in terms of the following paradigm~\cite{simmons}.  Alice and Bob are imprisoned in two different cells that are far apart.  They would like to devise an escape plan but the only way they can communicate with each other is through a courier who is loyal to and under the command of the warden (Eve, the adversary) of the penitentiary.  The courier leaks all information to the warden.  If the warden suspects that either Alice or Bob are conspiring to escape from the penitentiary, she will cut off all communication between them, and move both of them to a maximum security cell.  It is assumed that prior to their incarceration Alice and Bob had access to a shared secret key which they will later exploit to send secret messages hidden in a cover text.  Can Alice and Bob devise an escape plan without arousing the suspicion of the warden? 

Steganography is inherently different from cryptography.  In cryptography, an encryption algorithm acts on the secret message, utilizing a key (private or public).  The resulting message (or ``ciphertext''') looks like gibberish to an eavesdropper (Eve) who does not have access to the secret key.  However, by observing that the transmitted ciphertext is gibberish, Eve can conclude that the message contains private information.  By contrast, in steganography we do not necessarily need to encrypt the secret message.  Rather, we hide it within a larger plaintext message, often referred to as the ``covertext'' or ``cover-work.''  The resulting message, or ``stego-text,'' appears to be an irrelevant or benign message to Eve.  In the usual steganographic protocol we assume that Alice and Bob have access to a secure, shared secret key.  Alice uses this key together with an embedding function to hide the secret message, and Bob uses the same key to extract it.  Alice can go a step further by first encrypting the message and then hiding the ciphertext inside the covertext, generally at the cost of greater key usage.   Bob, on receiving the stego-text would first extract the ciphertext and then decrypt it to obtain the final secret message, using the shared key for both steps.    

Relatively little research has been done in quantum steganography.  The idea of hiding secret messages as the error syndromes of a quantum error-correcting code (QECC) was introduced by Julio Gea-Banacloche in~\cite{gea-banacloche:4531}.  In his formulation Alice and Bob use the three-bit repetition code to transmit classical messages to each other using a shared secret key.  All the noise in the channel that Eve perceives is because of these deliberate errors that Alice applies.  In his model he assumes that Alice and Bob share a binary-symmetric channel.  This work does not address the issue of whether the errors would resemble a plausible channel, nor does it consider the case where the channel contains intrinsic noise.  Natori gives a simple treatment of quantum steganography which is a modification of super-dense coding~\cite{natori}.  Martin introduced a notion of quantum steganographic communication based on a variation of Bennett and Brassard's quantum-key distribution (QKD), hiding a steganographic channel in the QKD protocol~\cite{martin}.  Curty et. al. proposed three different quantum steganographic protocols~\cite{curtysantos}.

In our earlier work \cite{shawandbrun}, we built on the approach of Gea-Banacloche in hiding information as errors in a quantum codeword.  We showed that such schemes can hide quantum as well as classical information; we proposed a quantitative measure of ``innocence'' (or secrecy), and derived rates of secret communication for Alice and Bob under particular assumptions about Eve's knowledge of the channel, as well as key-consumption rates.  We also addressed the issue of communicating over a channel containing intrinsic noise.
 
The day may come when quantum networks are ubiquitous.  Steganographic techniques may be useful as a way of authenticating quantum communications in distributed quantum information processing where the sender and receiver don't control all intermediate nodes of the network; such uses of steganography for authentication are often called ``watermarking'' in the classical case.  Quantum steganography gives a different kind of cryptographic protocol that may allow both secret and secure communication of quantum and classical information over quantum channels.

In Section~\ref{sec:hidingqi} we detail a protocol to hide four qubits in the $[[5,1,3]]$ five-qubit code (the ``perfect code'').  This protocol works by hiding information in the codewords of the perfect code in such a way that to Eve they look like depolarizing errors.  In Section~\ref{sec:optimize} we numerically estimate the optimal number of qubits that Alice can send to Bob on average using eight distinct encodings, and list the encoding using single-qubit errors.  We calculate the key consumption rates in Section~\ref{sec:keyconsume}.   We then consider encoding into a longer sequence of five-qubit code blocks in Section~\ref{sec:asymptotics}, and compare this to the theoretically achievable rate for a large block code.  We show that encoding across multiple blocks significantly enhances the secret communication rate, though it cannot quite match that of a large single code block.  We conclude with a discussion of these results.
%The Appendix lists all encodings in two-qubit errors.

%%%%%%%%%%%%%%%%%% NEW SECTION %%%%%%%%%%%%%%%%%%%%%%%%%%%%%%%%%%%%
\section{Hiding quantum information in the perfect code}
\label{sec:hidingqi}
\begin{figure}[ptb]
\begin{center}
\includegraphics[width=3.5in]{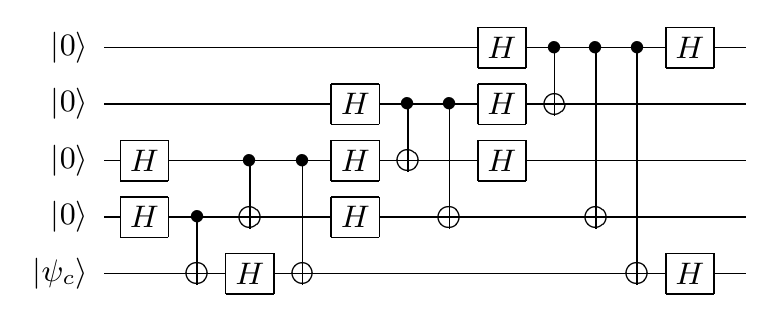}
\end{center}
\caption
{Encoding circuit for the perfect code.  The \textit{H} gate is a Hadamard gate, and the two-qubit gates are CNOT gates.  The first four qubits are ancilla qubits and the fifth qubit $\ket{\psi_{c}}$ is the cover-qubit.}
\label{fig:five_qubit_code}
\end{figure}

We introduced a general version of the protocol that we present in this section in Ref.~in~\cite{shawandbrun}, but this did not give examples of practical encoding techniques.  In the current paper, Alice uses the $[[5,1,3]]$ code (perfect code) to encode her stego-qubits in the syndromes of the code.  The perfect code encodes one logical qubit into five physical qubits, and is the smallest quantum error-correcting code that can correct an arbitrary single-qubit error~\cite{PhysRevLett.77.198}.  It is a nondegenerate code:  each single-qubit Pauli error is mapped to a unique syndrome.  The code is a stabilizer code with $n - k = 5 - 1 = 4$ stabilizer generators; we list them in Table~\ref{tbl:perfect_code}, along with the logical operators.  Alice and Bob can hide up to four qubits of information in this code, which they send over a channel that to Eve looks like a depolarizing channel.  Figure~\ref{fig:five_qubit_code} shows the encoding unitary circuit for the perfect code.

%ENCODING UNITARY CIRCUIT FOR THE PERFECT CODE

%TABLE FOR THE STABILIZER GENERATORS OF THE PERFECT CODE
\begin{table}[tbp] \centering
\begin{tabular}
[c]{c|ccccc}\hline\hline
$g_{1}$ & $X$ & $Z$ & $Z$ & $X$ & $I$\\
$g_{2}$ & $I$ & $X$ & $Z$ & $Z$ & $X$\\
$g_{3}$ & $X$ & $I$ & $X$ & $Z$ & $Z$\\
$g_{4}$ & $Z$ & $X$ & $I$ & $X$ & $Z$\\ \hline
$\overline{X}$ & $X$ & $X$ & $X$ & $X$ & $X$\\
$\overline{Z}$ & $Z$ & $Z$ & $Z$ & $Z$ & $Z$
\\\hline\hline
\end{tabular}
\caption{Stabilizer generators $g_1$, \ldots, $g_4$, and logical operators $\bar{X}$ and $\bar{Z}$ for the perfect code.  The convention in the above generators is that $Y=ZX$.}
\label{tbl:perfect_code}
\end{table}

The stabilizer generators in Table~\ref{tbl:perfect_code} can also be represented by a {\it symplectic matrix}, or $Z|X$ matrix~\cite{CleveGottesman97,grasslroetteler}.  This is the following binary matrix:
\begin{equation}
\left[  \left.
\begin{array}
[c]{ccccc}
0 & 1 & 1 & 0 & 0 \\
0 & 0 & 1 & 1 & 0 \\
0 & 0 & 0 & 1 & 1 \\
1 & 0 & 0 & 0 & 1 
\end{array}
\right\vert
\begin{array}
[c]{ccccc}
1 & 0 & 0 & 1 & 0 \\
0 & 1 & 0 & 0 & 1 \\
1 & 0 & 1 & 0 & 0 \\
0 & 1 & 0 & 1 & 0 
\end{array}
\right]
\label{perfectcodemat1}
\end{equation}
This matrix is formed from the stabilizer generators in Table~\ref{tbl:perfect_code}.  We obtain the the left $Z$ (left) submatrix by inserting a ``1'' wherever we see a $Z$ in the stabilizer generators.  We obtain the $X$ (right) submatrix by inserting a ``1'' wherever we see a corresponding $X$ in the stabilizer generator.  If there is a $Y$ in the generator, we insert a ``1'' in the corresponding row and column of both the $Z$ and $X$ submatrices.

We can use this matrix to find the error syndromes corresponding to single-qubit errors.  To generate the syndrome of each single-qubit error operator, we take the symplectic product of the parity check matrix~\ref{perfectcodemat1} with the binary vector of the error operator corresponding to a single-qubit error.  The perfect code has a total of fifteen single-qubit errors: $X_{1}, X_{2}, \ldots, X_{5}$, $Z_{1}, Z_{2}, \ldots, Z_{5}$ and $Y_{1}, Y_{2}, \ldots, Y_{5}$.  Together with the ``no error operator'' $IIIII$ these sixteen operators have sixteen distinct syndromes, which are listed in Table~\ref{tbl:syndrome_perfect_code}.  
%TABLE OF SYNDROMES FOR THE PERFECT CODE
\begin{table}[tbp] \centering
%EndExpansion%
\begin{tabular}
[c]{|c|c||c|c|}\hline\hline
Error   & Syndrome & Error & Syndrome\\ \hline \hline
$IIIII$ & $0000$ & $IIIIZ$ & $0001$ \\ \hline
$IIZII$ & $0010$ & $IIIXI$ & $0011$ \\ \hline
$IXIII$ & $0100$ & $IIXII$ & $0101$\\ \hline
$ZIIII$ & $0110$ & $IIYII$ & $0111$ \\ \hline
$XIIII$ & $1000$ & $IZIII$ & $1001$ \\ \hline
$IIIIX$ & $1010$ & $IIIIY$ & $1011$ \\ \hline
$IIIZI$ & $1100$ & $IYIII$ & $1101$ \\ \hline
$YIIII$ & $1110$ & $IIIYI$ & $1111$
\\\hline\hline
\end{tabular}
\caption{Syndrome table of the perfect code.  The error operators are ordered according to syndrome values.}
\label{tbl:syndrome_perfect_code}
\end{table}

\begin{figure}[htp]
  \begin{center}
    	 \includegraphics[width = 4.5in]{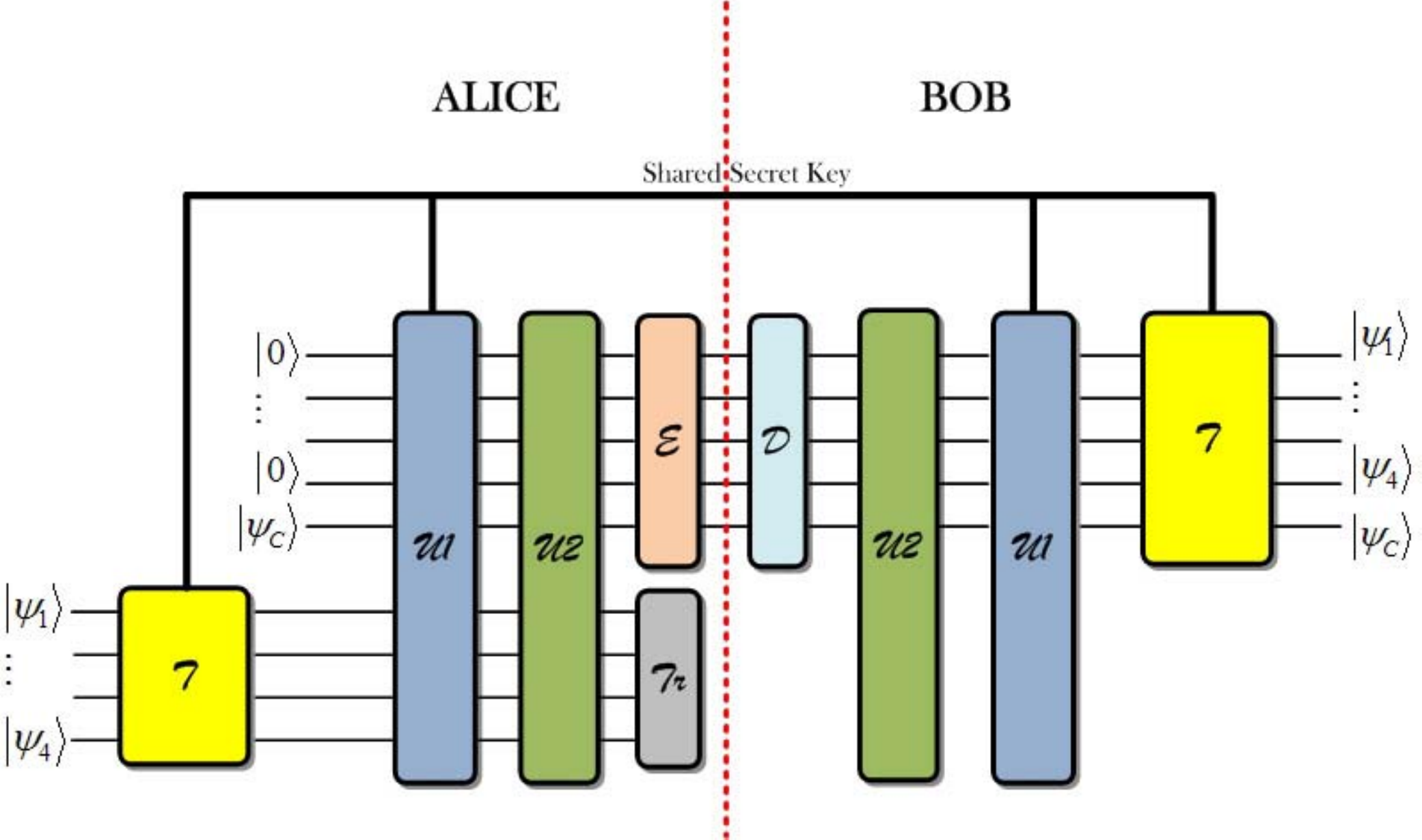}
  \end{center}
  \caption{(Color online) Steganographic protocol to hide four qubits of information $\ket{\psi_{1}},\ldots, \ket{\psi_{4}}$ using the perfect code.  Using a classical secret key shared with Bob, Alice first applies the twirling procedure (indicated in the figure by the yellow box) to each of her four qubits in the ``stego subsystem'' $S$.  This transforms each qubit into a maximally mixed state.  The codeword would be prepared using four ancilla qubits $\ket{0}$ along with a cover-qubit $\ket{\psi_{C}}$.  We call this ``subsystem $A$.''  The first unitary $\CU_{1}$ (blue box) is applied to the subsystems $A$ and $S$, followed by unitary $\CU_{2}$ (green box).  These have the effect of swapping the twirled stego qubits into the syndrome space of the code.  Finally Alice applies the encoding unitary $\CE$ to subsystem $A$ and sends it over a channel to Bob.  She traces out subsystem $S$ (gray box), which now is just in the state $\ket0$.  Upon receiving the codeword, Bob applies the decoding circuit $\CD$.  He undoes the unitary transformations $\CU_{1}$ and $\CU_{2}$ that were applied by Alice.  Finally, he uses the shared secret key to untwirl the qubits to obtain the original stego qubits.}
  \label{fig:stegoprotocol_perfect_code}
\end{figure}

We now detail the steps Alice takes to hide four stego qubits.  We label Alice's subsystem that contains the five-qubit codeword as $A$, and the subsystem holding the stego qubits as $S$.  Using the classical secret key she shares with Bob, Alice first applies a ``twirling'' operation to each of the four stego qubits.  That is, she randomly applies $I$, $X$, $Y$, or $Z$ to each qubit.  To do this to four qubits requires 8 bits of secret key.  (Note that if Alice's stego qubits are drawn from a source that looks like the maximally mixed state on average, twirling is not necessary.  For example, highly compressed quantum states will look close to maximally mixed.)  The combined $A$ and $S$ subsystems, after the twirling procedure but before encoding, are in the state:
\begin{equation}
\label{eqn:initialstate}
\rho^{SA}  =  \bigg(\frac{I}{2}\bigg)^{\otimes 4} \otimes  \ket{0000}\bra{0000} \otimes \ket{\psi_{C}}\bra{\psi_{C}}.
\end{equation}  
We can rewrite the $S$ subsystem of the state in Eq.~(\ref{eqn:initialstate}) in the computational basis as:
\begin{equation}
\label{eqn:Icomp}
\left(\frac{I}{2}\right)^{\otimes 4} = \frac{1}{16} \sum_{i = 0}^{15} \ket{k}\bra{k} ,
\end{equation}
where we denote the ket states from 0 to 15 in place of their binary representations.  We'll use the same notation for the four ancilla qubits in subystem $A$.  Combining Eq.~(\ref{eqn:Icomp}) with Eq.~(\ref{eqn:initialstate}) we get:
\begin{equation}
\label{eqn:initialstate2}
\rho^{SA} = \frac{1}{16}\sum_{k = 0}^{15}\ket{k}\bra{k}^{S} \otimes \ket{0}\bra{0}^{A}\otimes\ket{\psi_{C}}\bra{\psi_{C}}^{A},
\end{equation}
Alice now applies the unitary $U_{1}$ (shown in the blue box in Figure~\ref{fig:stegoprotocol_perfect_code}), defined as
\begin{equation}
\label{eqn:unitary1}
\CU_{1} \equiv \sum_{i = 0}^{15} \ket{i}\bra{i}^{S} \otimes \CE_{i}^{A} \otimes \CO_i^A ,
\end{equation}
where the $\CE_{i}$ and $\CO_i$ are the error operators listed in Table~\ref{tbl:encoded_errors_perfect_code}, ordered from top to bottom.  The operators $\CE_i\otimes\CO_i$ transform under the encoding unitary of the five-qubit code into the correctable one-qubit errors of Table~\ref{tbl:syndrome_perfect_code}.  Note that the $\CE_i$ operators act on the ancilla qubits so that  $\CE_i\ket0 \propto \ket{i}$.

\begin{figure}[ptb]
\begin{center}
\includegraphics[width=1.5in]{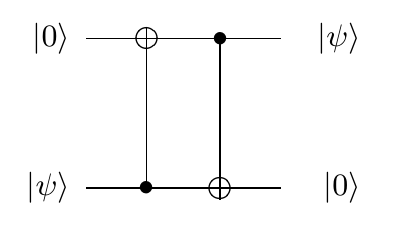}
\end{center}
\caption
{Swap circuit of a qubit with an ancilla in state $\ket0$.}
\label{fig:swap}
\end{figure}

We define the second unitary $U_{2}$ depicted by the green box in Figure~\ref{fig:stegoprotocol_perfect_code}, as:
\begin{equation}
\label{eqn:unitary2}
\CU_{2} \equiv \sum_{j = 0}^{15} (X^{j})^{S} \otimes \ket{j}\bra{j}^{A} \otimes I_{C}^{A}.
\end{equation}
In Eq.~(\ref{eqn:unitary2}) $X^{j}$ is a shorthand notation for $X^{j_{1}} \otimes X^{j_{2}} \otimes X^{j_{3}} \otimes X^{j_{4}}$, where $j_{1}\cdots j_{4}$ is the binary expression for $j$.  After applying the unitaries $\CU_{1}$ and $\CU_{2}$, the system is left in the state
\begin{equation}
\label{eqn:preencoded_state}
\rho' = \frac{1}{16}\sum_{k = 0}^{15}\ket{0}\bra{0} \otimes \ket{k}\bra{k}\otimes\CO_k\ket{\psi_{C}}\bra{\psi_{C}}\CO_k .
\end{equation}
Note that the original stego qubits are now in the state $\ket0$.  The quantum information they contained has been swapped into the syndrome subsystem.  The unitary operators $\CU_1$ and $\CU_2$ are analogous to the two CNOT operators in the usual circuit for swapping a qubit into an ancilla in the initial state $\ket0$ (see Figure~\ref{fig:swap}).

%TABLE OF ENCODED ERROR OPERATORS AND THE CORRESPONDING SYNDROMES
\begin{table}[tbp] \centering
%EndExpansion%
\begin{tabular}
[c]{|c|c||c||c|c|}\hline\hline
$\CE_i$   & $\CO_i$ & Encoded Error & Syndrome $i$ & Syndrome Label\\ \hline \hline
$IIII$ & $I$ & $IIIII$ & 0000 & $s_0$ \\ \hline
$IIIX$ & $Z$ & $IIIIZ$ & 0001 & $s_1$\\ \hline
$IIXI$ & $I$ & $IIZII$ & 0010 & $s_2$\\ \hline
$IIXX$ & $I$ & $IIIXI$ & 0011 & $s_3$\\ \hline
$IXII$ & $I$ & $IXIII$ & 0100 & $s_4$\\ \hline
$IXZY$ & $Y$ & $IIXII$ & 0101 & $s_5$\\ \hline
$ZXXI$ & $Z$ & $ZIIII$ & 0110 & $s_6$\\ \hline
$-IXYY$ & $Y$ & $IIYII$ & 0111 & $s_7$\\ \hline
$XIII$ & $I$ & $XIIII$ & 1000 & $s_8$\\ \hline
$XZIX$ & $I$ & $IZIII$ & 1001 & $s_9$\\ \hline
$XIXI$ & $X$ & $IIIIX$ & 1010 & $s_{10}$\\ \hline
$XIXX$ & $Y$ & $IIIIY$ & 1011 & $s_{11}$\\ \hline
$XXIZ$ & $X$ & $IIIZI$ & 1100 & $s_{12}$\\ \hline
$XYIX$ & $I$ & $IYIII$ & 1101 & $s_{13}$\\ \hline
$YXXI$ & $Z$ & $YIIII$ & 1110 & $s_{14}$\\ \hline
$XXXY$ & $X$ & $IIIYI$ & 1111 & $s_{15}$
\\\hline\hline
\end{tabular}
\caption{Encoded error operators of the perfect code.}
\label{tbl:encoded_errors_perfect_code}
\end{table} 

Alice then applies the encoding unitary $\CU_E$ to subsystem $A$, and traces out subsystem $S$.  The resulting codeword must look plausible to Eve.  In this case she is trying to match the depolarizing channel:
\begin{equation}
\label{eqn:depol_prefect_code}
\rho \rightarrow \CN(\rho) \equiv (1-p)\rho + \frac{p}{3}\sum_{i = 0}^{4} X_{i}\rho X_{i} + Y_{i}\rho Y_{i} + Z_{i}\rho Z_{i}.
\end{equation}
The correctly-encoded, error-free state would be
\begin{equation}
\rho_c = \CU_E \left( \ket{0000}\bra{0000} \otimes \ket{\psi_C}\bra{\psi_C} \right) \CU_E^\dagger.
\end{equation}
By swapping in the stego qubits into the syndrome space as described above, Alice has effectively prepared the state
\begin{equation}
\rho_c' = \sum_{k=0}^{15} E_k \rho_c E_k ,
\end{equation}
where the $E_k$ are the single-qubit errors from Table~\ref{tbl:syndrome_perfect_code} (including the ``no error'' operator $E_0=IIIII$).  To plausibly use the five-qubit code, the error rate $p$ must be sufficiently low that it is rare for more than one single-qubit error to occur.  So if Eve intercepts this state and checks the error syndromes, the result should look reasonable for a codeword that has passed through the depolarizing channel.

%%%%%%%%%%%%%%%% NEW SECTION %%%%%%%%%%%%%%%%%%%%%%%%%%%%%%%%%%%%%%%%%
\section{Average rate of steganographic transmission}
\label{sec:optimize}
While the encoding given in the previous section is reasonable for encoding stego-qubits in a single codeword, it would not do if Alice wants to send many messages to Bob.  It would quickly look suspicious if almost every codeword had exactly one single-qubit error.  For longer messages, Alice must most often apply no errors, from time to time apply a single error, and perhaps occasionally apply two errors (an uncorrectable error), in such a way as to match the statistics of a depolarizing channel with error probability $p$.

In this section we present details of how Alice optimizes the number of stego-qubits that she sends to Bob by encoding her stego-qubits in the syndromes of the perfect code.  She either sends no stego-qubits to Bob (and applies no errors to the codeword) or she sends four stego-qubits to him (applying either one or two errors).  We would like to maximize the number of stego-qubits that Alice can send to Bob under the constraint that Alice's encoding scheme should match the probability distribution of the depolarizing channel.  The channel model that we are interested in is:
\begin{widetext}
\begin{eqnarray}
\label{eqn:depolchannel}
\rho \rightarrow\CN(\rho) & \equiv & (1-p)^{5}\rho + \frac{p(1-p)^4}{3}\sum_{i = 0}^{4}\bigg(X_{i}\rho X_{i} + Y_{i}\rho Y_{i} + Z_{i}\rho Z_{i}\bigg) \nonumber \\
& + & \frac{p^2(1-p)^3}{9} \sum_{i<j} \bigg(X_{i}X_{j}\rho X_{i}X_{j} + X_{i}Y_{j}\rho X_{i}Y_{j} +  X_{i}Z_{j}\rho X_{i}Z_{j} \nonumber \\  
& + & Y_{i}X_{j}\rho Y_{i}X_{j} + Y_{i}Y_{j}\rho Y_{i}Y_{j} + Y_{i}Z_{j}\rho Y_{i}Z_{j} + Z_{i}X_{j}\rho 
+ Z_{i}X_{j} + Z_{i}Y_{j}\rho Z_{i}Y_{j} + Z_{i}Z_{j}\rho Z_{i}Z_{j}\bigg).
\end{eqnarray} 
\end{widetext}
There are of course three, four, and five qubit errors, but it is very unlikely that such errors will occur, and so we neglect them for this paper.  The techniques we give here could easily be used to find stego-encodings for those cases as well; but it might appear suspicious if Alice and Bob used an encoding that frequently suffered uncorrectable errors.

We use three distinct classes of encodings.  The first class corresponds to no errors, and transmits no hidden information.  This always has the syndrome $s_0 \equiv 0000$ in Table~\ref{tbl:encoded_errors_perfect_code}.

Alice's second encoding class corresponds to all single-qubit errors, plus no error, with equal probability.  These are the syndromes $s_{0} \equiv 0000$ to $s_{15} \equiv 1111$ in the same table.  This is the encoding described in the previous section.  So this encoding class includes a single encoding, and transmits four stego-qubits.

The third encoding class corresponds to the set of all two-qubit errors.  There are ninety such errors.  These ninety two-qubit errors naturally divide into six sets of fifteen errors each, where each set has one error corresponding to each of the fifteen nonzero syndromes.  By also including the ``no error'' operator, each set corresponds to an encoding that can transmit four stego-qubits.  We list these six sets in Table~\ref{tbl:two-qubit_errors_perfectcode}.  When Alice sends four stego-qubits to Bob, she must use all sixteen distinct syndromes.  A single row of this table, spanning all sixteen syndromes, corresponds to a single encoding.  Each encoding proceeds exactly as described above for the single-error case, except that the operators $\CE_i$ and $\CO_i$ from Table~\ref{tbl:two-qubit_errors_perfectcode} are used in Eq.~(\ref{eqn:unitary1}) instead of those from Table~\ref{tbl:encoded_errors_perfect_code}.

We now need to solve for how often the three classes of encodings should be used to match the channel statistics.  Let $Q_0$ be the fraction of times Alice uses the (trivial) first encoding class; $Q_1$ the fraction of times that Alice uses the second (single-error) encoding class; and $Q_2$ the fraction of times that Alice uses one of the six two-error encodings (which should be used equally often).  The channel distribution constraints are as follows:
\begin{eqnarray}
\label{eqn:channelconstraint1}
p_{0} & = & (1-p)^5 = Q_0 + \frac{1}{16}(Q_1 + Q_2) , \\
p_{1} & = & 5p(1-p)^4 = \frac{15}{16}Q_1, \\
p_{2} & = & 10p^2(1-p)^3 = \frac{15}{16}Q_2 . 
\end{eqnarray}
In the channel constraint equations above, $p_{0}, p_{1}$, and $p_{2}$ represent the total probability of the channel applying no errors, one error, or two errors on the codewords.  The right-hand-side of the equations represent how Alice matches the channel's probability distribution.  For example, she always applies no errors if she uses the first (trivial) encoding, and with probability $1/16$ if she uses one of the other encodings.  We solve for $Q_{0,1,2}$ to get:
\begin{eqnarray}
Q_0 &=& p_0 - (p_1+p_2)/15 , \nonumber\\
Q_1 &=& (16/15)p_1 , \\
Q_2 &=& (16/15)p_2 . \nonumber
\end{eqnarray}
Note that these numbers do not add up to 1, because we have neglected errors of weight three or more.  For small $p$ they will come close.

\begin{figure}[htp]
  \begin{center}
    	 \includegraphics[width = 3.5in]{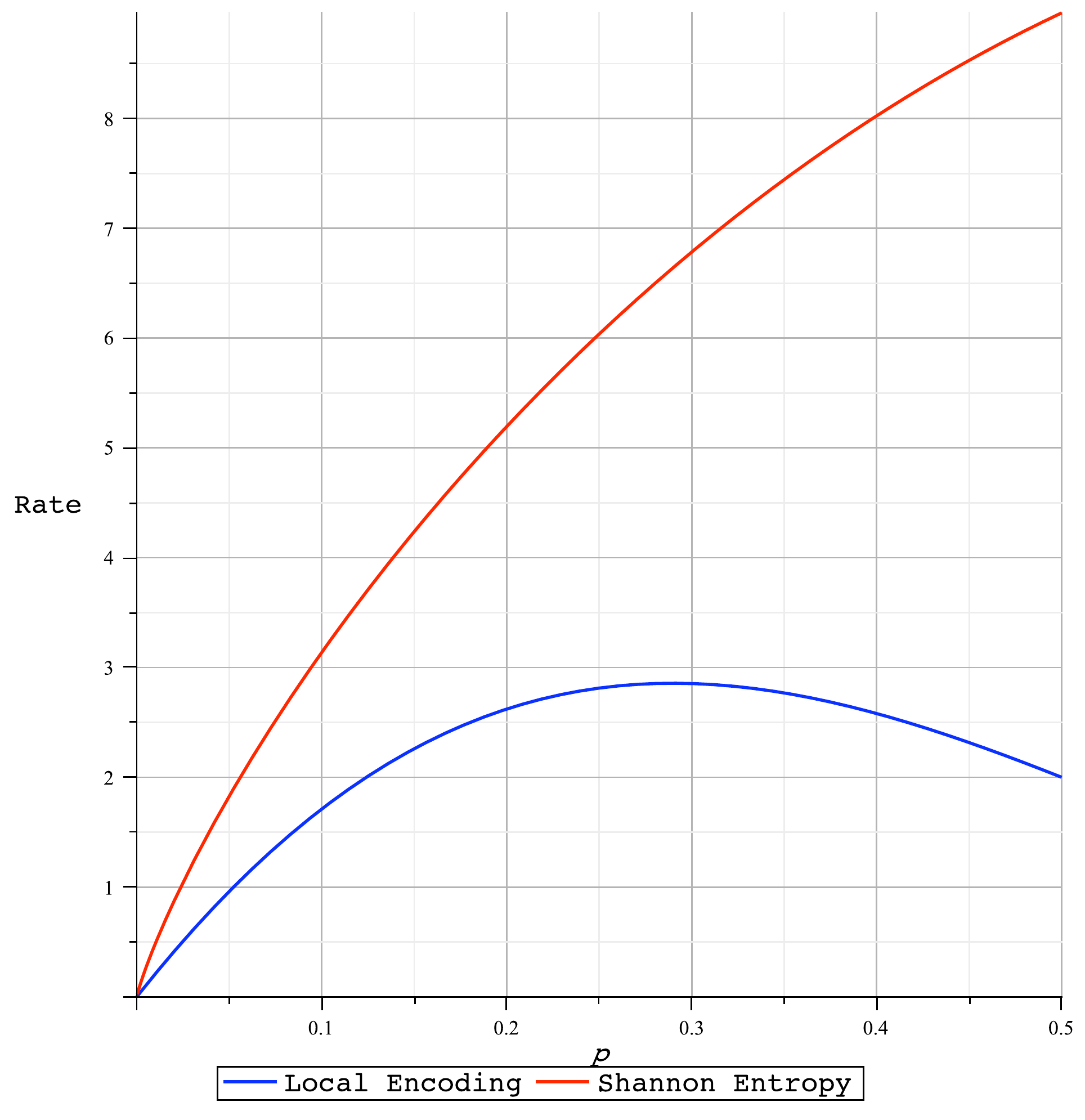}
  \end{center}
  \caption{(Color online) The red curve shows the Shannon entropy of the binary-symmetric channel.  The blue curve underneath the entropy curve is the rate of steganographic information maximized for various values of the error-rate $p$ of the channel.  In the limit of large $N$, encoding in a single large block (rather than multiple sub-blocks) can achieve a rate of steganographic transmission approaching the Shannon entropy \cite{shawandbrun}.}%  For  small $p$ one doesn't do too much worse using multiple finite-sized code blocks.}
  \label{fig:optimize}
\end{figure}

The average number of stego-qubits that Alice can send to Bob under the above constraints is $N_{avg} = 4(Q_1+Q_2) = (64/15)(p_1+p_2)$.  We plot this function in Figure~\ref{fig:optimize}, along with the Shannon entropy of the channel (which is the maximum possible rate at which stego information could be sent).  Note that this curve only makes sense for fairly small values of $p$; for higher values, it is no longer correct to neglect higher-weight errors.

%\section{Appendix}
%\label{sec:appendix}
%%%%%%%%%%% TABLE%%%%%%%%%%%%%%%%%%%%%%%%%%%%%%%%%%%%%%%%%%%%
\begin{table*}[tbp]\centering
\begin{tabular}[c]{|c|c|c||c|c|c|}
\hline \hline
Syndrome $i$ & $\CE_i\otimes\CO_i$ & Encoded Error & Syndrome $i$ & $\CE_i\otimes\CO_i$ & Encoded Error \\ \hline \hline
\multirow{6}{*}{0001} & $IZIXI$ & $XZIII$ & \multirow{6}{*}{0010} & $ZIXIZ$ & $XZIII$ \\ 
&  $\textbf{IIZYY}$ & $\textbf{IXXII}$ && $\textbf{IZXZX}$ & $\textbf{IYIYI}$ \\  
& $-ZIZYX$ & $ZIYII$ && $ZIXZY$ & $YIIZI$ \\
&  $IIIXI$ & $IIZXI$ && $IIXIX$ & $XIIIX$ \\
&  $ZIIYY$ & $YIIYI$ && $IZXIY$ & $IZIIY$	\\
& $-IZIYX$ & $IYIZI$ && $IIXIZ$ & $IIIXZ$ \\ \hline \hline
\multirow{6}{*}{0011} & $ZZXXZ$ & $YYIII$ & \multirow{6}{*}{0100} & $ZXIIZ$ & $ZIZII$ \\
& $\textbf{-ZIYYX}$ & $\textbf{ZIXII}$ && $\textbf{IXZZY}$ & $\textbf{IIYXI}$ \\ 
& $-IIYYY$ & $IXYII$ && $IXIZX$ & $XIIZI$ \\
&  $IZXXX$ & $IZIIX$ && $ZXIIY$ & $YIIIX$ \\
&  $IIXXY$ & $XIIIY$ && $IXIZZ$ & $IIIYY$	\\
&  $IIXXZ$ & $IIZIZ$ && $IXZZX$ & $IIXIZ$ \\ \hline \hline
\multirow{6}{*}{0101} & $IYIXI$ & $XYIII$ & \multirow{6}{*}{0110} & $IXXII$ & $IXZII$ \\
&  $\textbf{ZXIXZ}$ & $\textbf{ZIIXI}$ && $\textbf{IXYZY}$ & $\textbf{IIXXI}$ \\ 
& $IYIYX$ & $IZIZI$ && $-IYXZX$ & $IZIYI$ \\
&  $IXIYI$ & $IIIYX$ &&  $IXXZI$ & $IIIZX$ \\
&  $-ZXIXX$ & $YIIIY$ &&  $IYXIY$ & $IYIIY$	\\
& $IXIXZ$ & $IXIIZ$ && $-IXYZX$ & $IIYIZ$ \\ \hline \hline
\multirow{6}{*}{0111} & $-ZYXXZ$ & $YZIII$ & \multirow{6}{*}{1000} & $XZZZY$ & $IYXII$ \\
&  $\textbf{IXXXI}$ & $\textbf{IXIXI}$ && $\textbf{XIZIZ}$ & $\textbf{IIYYI}$ \\ 
&  $IXXYX$ & $XIIYI$ &&  $XIIZX$ & $IXIZI$ \\
&  $IYXXX$ & $IYIIX$ &&  $XIIIX$ & $IIZIX$ \\
& $-IXXYZ$ & $IIIZY$ &&  $XIIIY$ & $IIIXY$	\\
&  $ZXXXI$ & $ZIIIZ$ &&  $XZIIZ$ & $IZIIZ$ \\ \hline \hline
\multirow{6}{*}{1001} & $-YIZYX$ & $YIYII$ & \multirow{6}{*}{1010} & $YIXIZ$ & $YXIII$ \\
& $\textbf{-YIIYY}$ & $\textbf{ZIIYI}$ && $\textbf{-XZYZY}$ & $\textbf{IYYII}$ \\ 
&  $XIZXZ$ & $IIXZI$ &&  $XIXII$ & $XIZII$ \\
&  $XIIXX$ & $IIIXX$ &&  $XZXII$ & $IZIXI$ \\
&  $XIIXY$ & $IIZIY$ && $XIYIZ$ & $IIXYI$	\\
&  $XIIXZ$ & $XIIIZ$ && $-YIXZY$ & $ZIIZI$ \\ \hline \hline
\multirow{6}{*}{1011} & $-YZXXZ$ & $ZYIII$ & \multirow{6}{*}{1100} & $XXIII$ & $XXIII$ \\
& $\textbf{-YIYYX}$ & $\textbf{YIXII}$ && $\textbf{-XYZZY}$ & $\textbf{IZXII}$ \\ 
&  $XZXXI$ & $IZZII$ && $YXIIZ$ & $YIZII$ \\
&  $XIXXI$ & $XIIXI$ && $-YXIIY$ & $ZIIIX$ \\
&  $XIXYX$ & $IXIYI$ && $XXZZI$ & $IIYIY$	\\
& $-XIYXZ$ & $IIYZI$ && $XYIIZ$ & $IYIIZ$ \\ \hline \hline
\multirow{6}{*}{1101} & $XXZYY$ & $XIXII$ & \multirow{6}{*}{1110} & $XYYZY$ & $IZYII$ \\
&  $\textbf{YXIXZ}$ & $\textbf{YIIXI}$ && $\textbf{XYXII}$ & $\textbf{IYIXI}$ \\ 
&  $XXIYX$ & $IIZYI$ &&  $XXXZX$ & $IIZZI$ \\
&  $XXZYZ$ & $IIYIX$ && $XXXIX$ & $IXIIX$ \\
&  $YXIXX$ & $ZIIIY$ && $XXYZI$ & $IIXIY$	\\
&  $XXIYY$ & $IIIZZ$ && $-XXXZY$ & $IIIYZ$ \\ \hline \hline
\multirow{6}{*}{1111} & $YYXXZ$ & $ZZIII$ & \multirow{6}{*}{0000} & $IIIII$ & $IIIII$ \\
& $\textbf{-XXYYY}$	 & $\textbf{XIYII}$ && $\textbf{IIIII}$ & $\textbf{IIIII}$  \\	
& $XYXXI$	 & $IYZII$ && $IIIII$ & $IIIII$ \\	
& $XXYYZ$	 & $IIXIX$ && $IIIII$ & $IIIII$ 	\\
& $XXXXY$	 & $IXIIY$ && $IIIII$ & $IIIII$ 	\\
& $YXXXI$	 & $YIIIZ$ && $IIIII$ & $IIIII$ 	\\ \hline \hline
\end{tabular}  
\caption{Table of double-qubit errors.  The error operators in bold represent a single encoding.  The table represents a total of six different encodings where each encoding utilizes sixteen distinct error operators and their corresponding distinct syndromes.}
\label{tbl:two-qubit_errors_perfectcode}
\end{table*}

%%%%%%%%%%%%%%%%%% NEW SECTION %%%%%%%%%%%%%%%%%%%%%%%%%%%%%%%%%%%%%%%%%
\section{Key Consumption}
\label{sec:keyconsume}
We define the key consumption rate as the number of classical bits of key consumed by Alice and Bob per qubit block.  Alice can send an $N$-qubit block to Bob by combining $N/5$ five-qubit blocks together.  How much shared secret key do Alice and Bob require to send steganographic information to each other?  In Section~\ref{sec:optimize} we defined $Q_{0}$ as the probability of Alice using the first encoding to send no stego-qubits to Bob.  The probability of using the second encoding is $Q_{1}$, when Alice applies single errors to her codewords.  Alice and Bob can also use two-error encodings to send four qubits to each other, of which there are a total of six as shown in Table~\ref{tbl:two-qubit_errors_perfectcode}.  We assume that each of these six encodings are equiprobable, each with probability $Q_2/6$.  Therefore, there are a total of eight different encodings that Alice and Bob must choose from in order to send steganographic information to each other.

For a single five-qubit block, Alice and Bob must share three classical bits to specify which of the eight encodings has been used.  They also require an extra eight bits for Alice to construct her four twirled qubits and for Bob to untwirl these qubits once he receives the steganographic information.  So Alice and Bob would consume twelve classical secret key bits for each five-qubit block, if they encode one block at a time, and so for an $N$-qubit block they would consume at most $12N$ classical bits of key---not a very economical protocol. In fact, it would be less than that, since Alice needs no key bits to do twirling for the first, trivial, encoding.  Taking this into account, the total key consumption rate would be $(3+8(Q_1+Q_2))/5$ secret key bits per qubit.

However, Alice and Bob can do much better than this by encoding into multiple block at once.  In the limit of large $N$, the number of bits that they require to specify the key is given by the Shannon entropy of the probability distribution of the three different encodings.  This gives us a key usage rate of
\begin{equation}
\label{eqn:keyconsume}          
K =  \frac{1}{5} \left(- Q_{0}\log_{2}{Q_{0}} - Q_{1}\log_{2}{Q_{1}} - Q_2\log_{2}{Q_2/6}\right)
\end{equation}
key bits per qubit.  For small $p$ this dramatically outperforms the rate for encoding one block at a time.
\\

%%%%%%%%%% NEW SECTION %%%%%%%%%%%%%%%%%%%%%%%%%%%%%
\section{Asymptotics}
\label{sec:asymptotics}
\begin{figure}[htp]
  \begin{center}
    	 \includegraphics[width = 3.5in]{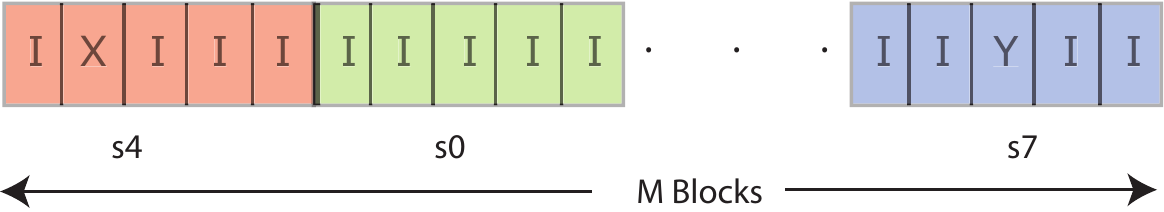}
  \end{center}
  \caption{(Color online) Alice sends a total of $M$ blocks consisting of $N = 5M$ qubits.  We label each block with its corresponding syndrome, $a_{0}, a_{1}, \textellipsis , a_{M}$.  Each $a_{j}$ corresponds to a syndrome from Table~\ref{tbl:encoded_errors_perfect_code}.  To obtain the syndrome for the $N$-qubit block we simply concatenate the syndromes.}
  \label{fig:fivequbitblockcode}
\end{figure}

We would like to quantify the steganographic rate when Alice sends a large block code to Bob.  We depict the asymptotic scenario in Figure~\ref{fig:fivequbitblockcode}.  The channel has no intrinsic noise of its own.  Whatever noise that Eve observes on the channel is due to Alice.  We address the problem of where the channel has intrinsic noise in~\cite{shawandbrun}.The $N$-qubit block code is constructed by concatenating $M$ five-qubit blocks.  Alice hides her steganographic information in the $N$-qubit block, and we would like to estimate the number of stego-qubits per $N$-qubit block that Alice can send to Bob while matching the statistics of the depolarizing channel.  Because of the low error-rate of the channel most of the time Alice applies no error to her $M$ code blocks.  This corresponds to syndrome $s_{0}$ from Table~\ref{tbl:encoded_errors_perfect_code}.  Occasionally she applies a single error.  We label each five-qubit block by $a_{j}$, where $j \in \{1, 2, \textellipsis , M\}$.  Each $a_{j}$ corresponds to one of the syndromes from Table~\ref{tbl:encoded_errors_perfect_code}.  In this section we define a string to be a string of syndromes for a concatenated $M$ five-qubit blocks where each five-qubit block is labelled by its corresponding syndrome.  

\begin{figure}[htp]
  \begin{center}
    	 \includegraphics[width = 3.5in]{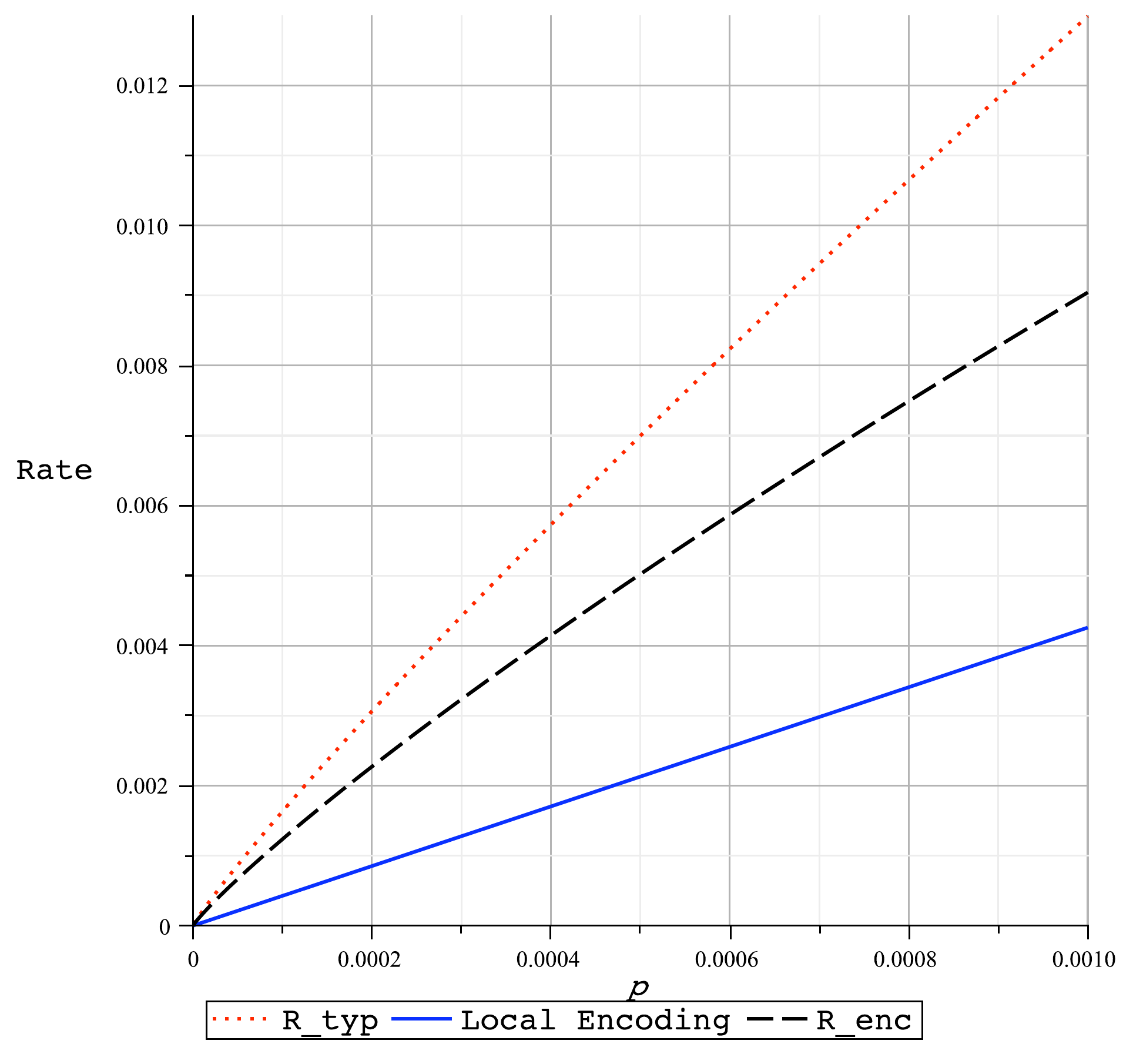}
  \end{center}
  \caption{(Color online) The green curve for $R_{typ}$ outperforms the rates from the syndrome encoding (black) and the local encoding (blue).  To generate the black curve we chose $\delta = 0.005$.}
  \label{fig:rate_plot}
\end{figure}

Let $p_{0} = (1-p)^{5}$ be the probability of no error on a block.  Let $p_{1} = 1-(1-p)^{5}$ be the probability of a single error on a block.  For small values $p$ we can approximate $p_{0} = (1-5p)$ and $p_{1} = 5p$.  If Alice hides information in typical sequences then the total number of such sequences is approximately $2^{MH(p)}$, where $H(p)$ is the Shannon entropy of the channel.  So the number of qubits that Alice can hide with this encoding will be approximately $MH(p)$.  We first calculate the Shannon entropy $H(p)$:
\begin{eqnarray}
H(p) & \approx & -(1-p)^{5}\log_{2}(1-p)^{5} \\ \nonumber
& - & 15\frac{1-(1-p)^{5}}{15}\log_{2}\bigg(\frac{1-(1-p)^{5}}{15}\bigg)~, \\ \nonumber
& \approx & -(1-5p)\log_{2}(1-5p) - 5p\log_{2}5p + 5p\log_{2}15~. \nonumber
\end{eqnarray}
The above is an approximation because we ignore the contribution to the Shannon entropy from double and higher errors.  We define the steganographic rate as the number of steganographic qubits that Alice sends to Bob per $N$-qubit block. In our case $N = 5M$.  So the rate $R_{typ} = \log_{2}2^{MH(p)}/5M$.  We can expand this further and write:
\begin{equation}
\label{eqn:Rtyp}
R_{typ} \approx -\frac{1}{5}\log_{2}(1-5p) + p\log_{2}\bigg(\frac{3(1-5p)}{p}\bigg)~.
\end{equation} 
The total probability $p_{k}$ of all strings of weight $k$ is:
\begin{eqnarray}
\label{eqn:totprob}
p_{k} & \approx &{M \choose k}\bigg(\frac{1-(1-p)^{5}}{15}\bigg)^{k}(1-p)^{5(M-k)}15^{k}~, \\ \nonumber
& \approx & {M \choose k} \bigg(\frac{p}{3}\bigg)^{k}(1-5p)^{M-k}15^{k}~. \nonumber
\end{eqnarray}
The average number of errors in an $N$-qubit code block goes like $5Mp$.  Let the variance be $5Mp\delta$, where $0 < \delta \ll 1$.  For each $k$ from $5pM(1-\delta)$ to $5pM(1+\delta)$, Alice can choose $C_{k}$ strings of weight $k$.  Let
\begin{equation}
\label{eqn:C}
C = \sum_{k = 5Mp(1-\delta)}^{5Mp(1+\delta)} C_{k}~.
\end{equation}
%%%%% FIGURE %%%%%%%%%%%%%%%%
Define the probability $q = 1/C$.  Then we would like to satisfy:
\begin{equation}
qC_{k} = C_{k}/C \approx p_{k}~.
\end{equation}
The choice for the number of strings of weight $k$ cannot exceed ${M \choose k}$, and so $C_{k} \leq {M \choose k}$.  From this constraint we obtain
\begin{equation}
\label{eqn:qconstraint}
q \geq \bigg(\frac{p}{3}\bigg)^{k}(1-5p)^{M-k}15^{k}.
\end{equation}
Alice would like C to be as large as possible which means that $q$ must be as small as possible.  This further implies that we should set $k = 5pM(1-\delta)$.  This give us
\begin{equation}
\label{eqn:Cequation}
C \approx (p/3)^{-5Mp(1-\delta)}(1-5p)^{5Mp(1-\delta)-M}15^{-5Mp(1-\delta)}~.
\end{equation}
So the steganographic rate for this encoding $R_{enc} \approx \log_{2}C/5M$.  Finally,
\begin{equation}
R_{enc} \approx -\frac{1}{5}\log_{2}(1-5p) + p\log_{2}\bigg(\frac{(1-5p)(5p)^{\delta}}{5p(1-5p)^{\delta}}\bigg)~.
\end{equation}
We plot the local encoding from Section~\ref{sec:optimize}, along with $R_{typ}$ and $R_{enc}$ for low values of $p$ in Figure~\ref{fig:rate_plot}.  As expected the typical sequences encoding outperforms the local encoding as well as the syndrome encoding described above.  However, we do much better asymptotically with the current encoding as opposed to the local encoding.   

%%%%%%%%%%%%%%% NEW SECTION %%%%%%%%%%%%%%%%%%%%%%%%%%%%
\section{Summary and Conclusion}

Steganography is the art of hiding information by disguising it as something else.  This idea, developed in the classical sphere, can be adapted to hide quantum as well as classical information---in this case, by making the quantum information appear to be errors from a quantum depolarizing channel.  Alice and Bob conceal their communication from Eve, hiding their message as errors in a codeword for an ``innocent'' cover state $\ket{\psi_c}$, and using the resource of a shared secret random key.  (Shared entanglement would work as well, or even better.)

In this paper, we have made use of the ideas from our earlier work \cite{shawandbrun} for a particular finite code:  the five-qubit ``perfect'' code.  We have explicitly presented the encodings needed to simulate the quantum depolarizing channel, and calculated the rate of transmission of steganographic qubits and the rate of usage of the shared secret key.  While we only worked out encodings that appear to be single-qubit or two-qubit errors, the same techniques of this paper will readily yield encodings for higher-weight errors as well.

All of this work assumed that the underlying physical channel used by Alice and Bob is error-free.  It merely appears noisy to Eve because of the machinations of Alice.  In principle, steganographic information can still be hidden even if the underlying channel is noisy.  This requires an additional layer of error correction, to protect the hidden information from physical noise.  In practice, these encodings are not necessarily straightforward, since errors that are local in physical space may appear quite differently in the space of syndromes of the code for the covertext.  The details of such encodings, and what rates can be practically achieved, remain work for the future.

\section*{Acknowledgments}
BAS and TAB would like to thank Daniel Gottesman, Patrick Hayden, Debbie Leung, Daniel Lidar, John Preskill, Stephanie Wehner and Mark Wilde for their useful comments and suggestions at various stages of this work.  This work was funded in part by NSF Grant No.~CCF-0448658.  TAB also acknowledges the hospitality and support of the Kavli Institute for Theoretical Physics in Santa Barbara.
%%%%%%%%%%%%%%%%%%%%%%%%%%%%%%%%%%%%%%%%%%%%%%%%%%%%%%%%%%%%%%%%%%%%%%%%%%%%%%%%%%%%%%%
%%%%%%%%%%%%%%%%%%%%%%%%%%% BIBLIOGRAPHY %%%%%%%%%%%%%%%%%%%%%%%%%%%%%%%%%%%%%%%%%%%%%%

%\bibliographystyle{apsrev}
\bibliography{stegoexample}

\end{document}